\documentclass[runningheads]{llncs}

\usepackage{tabularx}
\usepackage{booktabs}
\usepackage[T1]{fontenc}
\usepackage{graphicx}
	\usepackage{mdframed}
\mdfdefinestyle{style1}{leftmargin=1cm,rightmargin=0.5cm,skipbelow=0.5cm,skipabove=0.2cm,
   innertopmargin=2pt}
   
\begin{document}
\title{(Over)Reliance on Test Agents in\\ AI-Assisted Software Testing}
%
%
\author{Eduard Paul Enoiu \orcidID{0000-0003-2416-4205}}
\authorrunning{E.P. Enoiu}
%
\institute{Department of Computer Science and Engineering\\
Mälardalen University, Västerås, Sweden
\\
\email{eduard.paul.enoiu@mdu.se}}
\maketitle              
\begin{abstract}
AI-based test agents promise to accelerate software testing by shortening feedback loops in continuous development and improving scalability and maintainability. To realize these benefits, engineers must still be able to assess if agent outputs are useful, valid, and reliable, rather than treating them as credible because they come from a capable system. This paper argues that overreliance on AI in testing is both an agency problem, in which engineers may cede cognitive control over test design decisions, and an assurance problem, in which testing artifacts may be accepted as evidence without sufficient scrutiny. We develop this argument through three theoretical lenses: software testing as cognitive problem-solving, test agents as adaptively autonomous entities, and test design argumentation as a means of making generated tests reviewable. We propose a framework for collecting data on overreliance in test agent workflows and identify specific modes of overdependence. The goal is to support accelerated testing without weakening judgment or the assurance value of testing evidence.

\end{abstract}
\section{Introduction}
Modern AI-assisted software development produces rapid code changes and large volumes of test artifacts, increasing the cognitive load on software engineers \cite{fakhoury2024llm,chen2025agenttester,harman2025mutation}. This scale and pace make it difficult to maintain oversight, assess quality, and make sound decisions, even as expectations for reliability remain high \cite{strandberg2025ethical}. The current challenge is to ensure that testing supports human understanding and judgment when AI agents are used during software testing. Prior work introduced the notion of \textit{test agents} and adaptive autonomy in regression test selection \cite{enoiu2019test,kumaresen2020agent}, arguing that test selection and scheduling decisions can benefit from decentralized control. A recent study on ethical challenges in software test automation \cite{strandberg2025ethical} identified explainability, logging and monitoring, privacy, technical risks, and human control as central concerns when AI is introduced into test automation. This raises a specific question for software testing: when LLMs explain generated tests, do engineers treat those explanations as hypotheses to inspect or as assurance to accept \cite{akbarova2026understanding}. 

Capabilities that make test agents useful also create risks. If a test agent can generate tests, explain why they exist, summarize execution results, delegate subtasks to other agents and regenerate itself after software changes, then the tester role shifts. The human becomes a supervisor for (semi)autonomous work, and overreliance on test agents can indirectly weaken human oversight by undermining the evidence that justifies confidence.

This paper makes the following contributions. It frames overreliance on test agents as a supervisory control and assurance problem and develops this framing through three theoretical lenses: software testing as cognitive problem-solving, test agents as adaptively autonomous test artifacts, and test design argumentation for making generated tests more reviewable. In the end, it proposes a categorization of overreliance modes specific to AI-assisted software testing and outlines implications for test agent workflows and empirical studies. 


\section{Related Work and Theoretical Lens}
Automated test generation techniques have been used in safety-critical development, such as in industrial control software \cite{enoiu2016automated}. Several studies \cite{kurmaku2022human} have shown that automated test generation is often more efficient than manual testing but remains less effective than manual test design by experienced engineers in detecting naturally occurring faults.

Research on the cognitive foundations of software testing \cite{enoiu2020towards,aniche2021developers,itkonen2012role} has focused on tester routines and problem-solving \cite{enoiu2023understanding}. Together, these results suggest that test design relies on goal formulation, selection of test design methods, and goal representation and context-based heuristics that are rarely captured in current automated approaches. This cognitive perspective is central to this work.

A review \cite{romeo2026exploring} found that automation bias is evident in testing practice \cite{enoiu2024unveiling}, making overreliance a particularly important focus. Biases such as confirmation bias and anchoring seem to be present in practice \cite{enoiu2024unveiling}, making overreliance especially important to study in test design and review. This framing builds on classic automation literature showing that automation can improve performance and reduce human involvement in monitoring, making intervention more difficult when problems occur~\cite{bainbridge1983ironies,parasuraman2000model}.

In safety-critical software development, tests can contribute to claims about system behavior and release readiness through verification evidence and assurance practices such as ISO 26262 \cite{griessnig2017development}, EN 50128 \cite{en200150128}, and the use of GSN \cite{spriggs2012gsn}.

\begin{figure}
   \centering
     \includegraphics[width=0.92\textwidth]{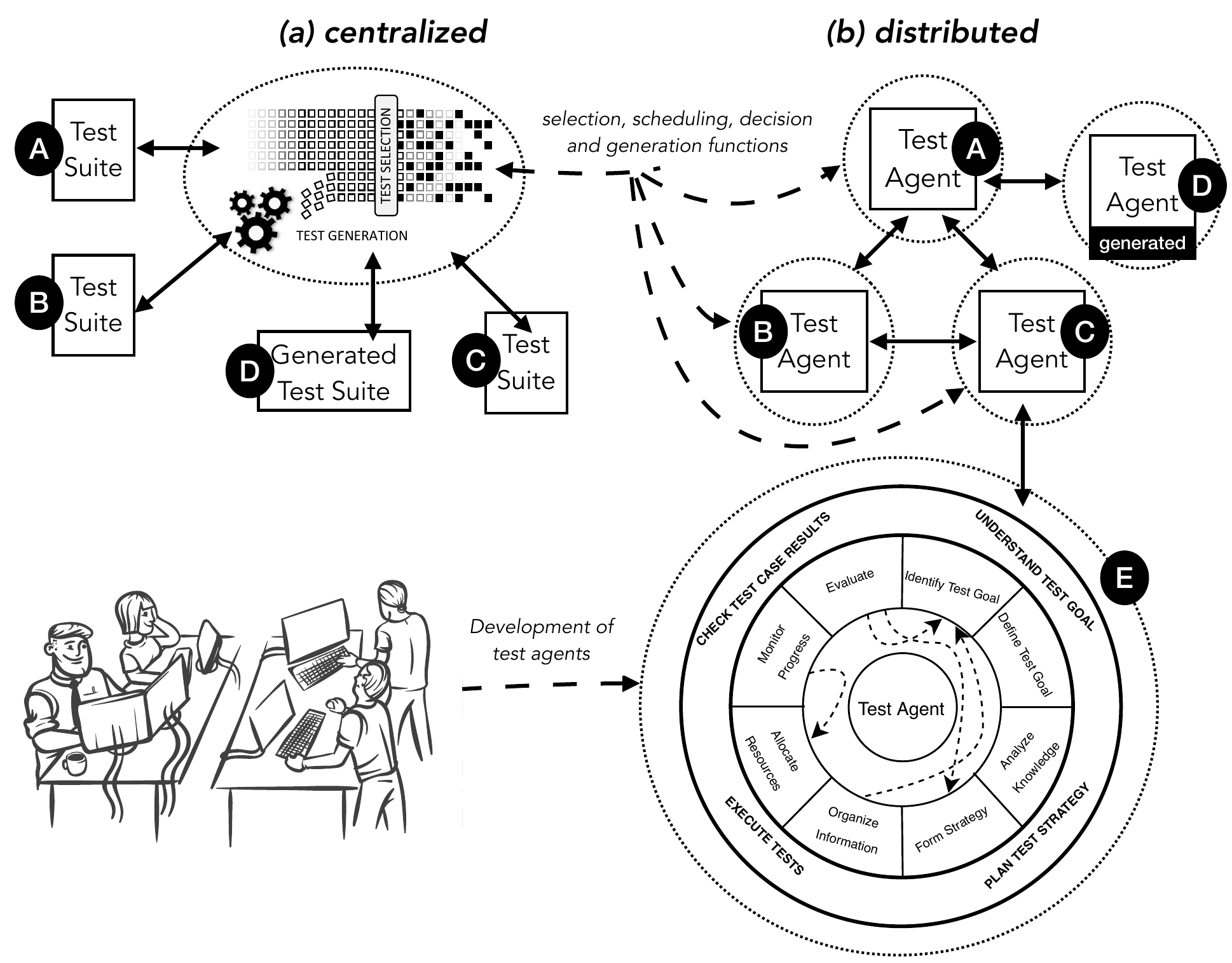}
        \vspace{-5mm}
   \caption{From centralized regression testing (a) to supervised test agent workflows (b). The internal agent problem-solving model is included to indicate which elements of agent behavior should remain inspectable for reliance.}\label{overall}
   \vspace{-2mm}
\end{figure}

\subsection{From Test Cases to Test Agents}

Building on the concept of adaptive autonomy for \textit{test agents} \cite{enoiu2019test,kumaresen2020agent}, one can view software test agents as capable of adjusting their autonomy levels in response to the testing context, available information, and observed system conditions. Instead of relying on a single centralized mechanism to decide which test cases to select or prioritize, one can rely on a decentralized form of coordination among agents. As illustrated in Figure~\ref{overall}, this allows individual agents A, B, C, or D to make local decisions, interact with other agents when needed, and adapt their behavior. Traditional automated test cases are typically static artifacts. They contain input values, expected results, and executable scripts. They are selected, scheduled, and prioritized by a surrounding regression test system. In contrast, the test-agent vision proposes test cases that can reason, adapt, interact, and update their behavior over time. Test agents are intended to decentralize regression testing by allowing tests to know when to execute, how to adapt their purpose, and when to interact with other tests. 

To avoid using the term \emph{test agent} as a loose term for any AI tooling, we adopt the following definition in line with prior work on test agents and agent-based software testing~\cite{enoiu2019test,kumaresen2020agent}.

\begin{mdframed}[linewidth=1pt, linecolor=black, roundcorner=5pt]
\small{\textbf{Definition (Test Agent).} A test agent is an autonomous or semi-autonomous software agent embedded in a testing workflow that can observe changes in the system under test and its testing environment, maintain explicit testing goals, generate, select or execute test artifacts, interact with other agents or humans and produce evidence relevant to assurance.}
\end{mdframed}

This changes what engineers must review. A test case becomes a partially autonomous entity in a testing workflow. Earlier test agent work \cite{enoiu2019test} describes agents with states such as Idle, Interact, Execute, Regenerate, and Out of Order. A test agent may execute its own task, request assistance, respond to another agent, or regenerate with help from a test engineer when its original purpose no longer holds. Test agent interactions can include non-committal information sharing, one-to-one dialogue, one-to-one delegation, and one-to-many dialogue or delegation, for example, when agents coordinate around coverage, execution time, or fault history.  This means that test agents operate across several layers of testing work. They may generate tests, evaluate results, exchange evidence, delegate goals, monitor changes, and change future test execution. In the language of semi-executable artifacts, such workflows combine executable code, prompts, agent workflows, evaluation harnesses, policies, and human judgment. Feldt et al.'s \cite{feldt2026agentic} semi-executable stack describes these as artifacts whose behavior depends partly on deterministic execution and partly on human interpretation.

\subsection{Cognitive Problem Solving and Argumentation}

To understand overreliance in testing, one must first understand what human testers do. Software testing is not only about producing test inputs or executing scripts \cite{ammann2016introduction}. It is a \textit{cognitive problem-solving activity} \cite{enoiu2023understanding}. Testers interpret requirements, infer risks, formulate test goals, select techniques, construct test cases, judge adequacy, evaluate results, and communicate their meaning to others (as shown inside test agent E in Figure~\ref{overall}). AI-assisted testing can change this cognitive structure. It can support or replace parts of the tester’s reasoning process. A model may propose the test goal, generate a concrete test, state the claim the test supports, provide a rationale, identify evidence, summarize the execution, and recommend whether a suite is adequate. For example, overreliance could occur when the engineer stops treating these outputs as hypotheses to be examined and starts treating them as evidence to be accepted. In complementary work on AI-assisted test generation, generated tests can be related to an explicit test goal, claim, reason, and evidence \cite{Enoiu7368} through \textit{test argumentation}. These results are relevant to overreliance, as they provide the initial conceptual basis for argumentative and inspectable AI-assisted testing.

\section{A Framework for Studying Overreliance on Test Agents}

This section turns the theoretical lenses into a conceptual framework for studying overreliance on test agents. It explains why the shift toward test agents makes overreliance both a supervisory-control and assurance problem, and it describes a data-collection setup for analyzing how test agents are implemented, used, evaluated, and revised. Finally, it identifies test-specific modes of overreliance that can serve as a checklist for designing test agent workflows.

\subsection{Overreliance as a Supervisory Control and Assurance Problem}

Prior work on ethical AI-powered test automation \cite{strandberg2025ethical} identifies human control and responsibility as central concerns, asking how humans remain in control, how they interact with automation, how inadequate performance is detected, and who is responsible when automation underperforms. The same work explicitly raises risks such as over-trust and flawed decision support in test automation. 

Test agents \cite{enoiu2019test} can amplify these risks. A statically generated test can be inspected. A test agent can act over time. It can change state, request help, delegate subtasks, regenerate, and produce artifacts and summaries. The human supervisor must therefore understand why it acted, which evidence it used, and when its results should be challenged.

Testing is also an assurance activity \cite{ammann2016introduction}. It supports claims about software behavior, quality, risk, and release readiness. In regulated and safety-critical settings, testing evidence often contributes to larger quality assurance. Prior work on test design argumentation \cite{Enoiu7368} notes that regulated domains require evidence that testing has met integrity and certification objectives, and that argumentation can structure claims, strategies, and supporting evidence, even for individual test-case design. In AI-assisted testing, overreliance could mean accepting agent-produced testing artifacts, arguments, summaries, delegations, or regenerated suites as assurance without sufficient challenge of their arguments.

\subsection{Data Collection for Overreliance on Test Agents}
\begin{figure}[t]
   \centering
     \includegraphics[width=0.80\textwidth]{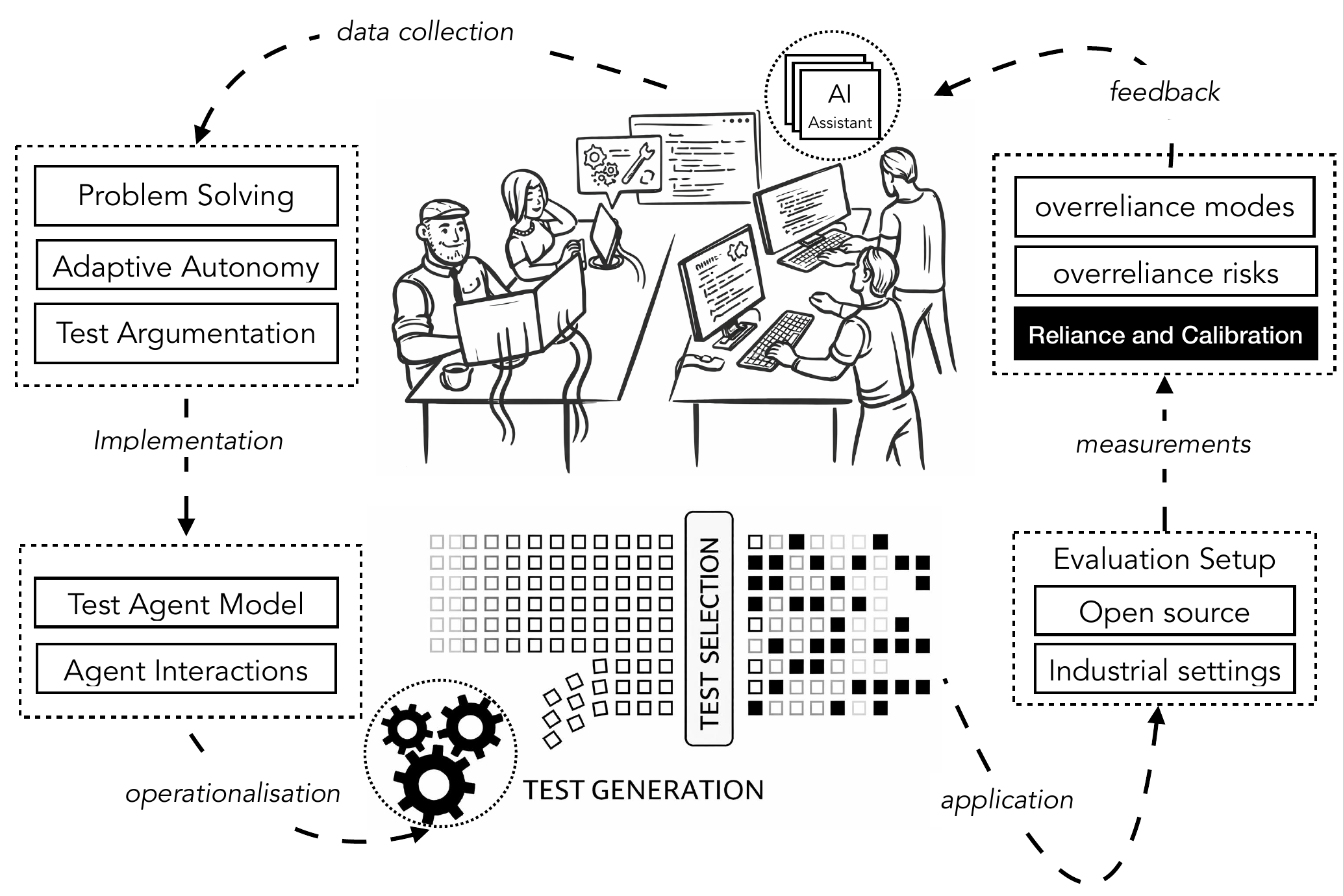}
        \vspace{-2mm}
   \caption{Conceptual framework for studying overreliance-aware test agents. The framework links theoretical lenses, implementation of adaptive and argumentative test agents, and empirical evaluation in open-source and industrial settings.}\label{fig:proj}
\end{figure}

Figure \ref{fig:proj} summarizes the steps proposed for future data collection. Building on the concept of adaptive autonomy for test agents \cite{enoiu2019test,kumaresen2020agent}, software test agents can adjust their level of autonomy based on the testing context, available information, and observed system conditions. Cognitive models of test design and adaptive test agents \cite{enoiu2019test,enoiu2020towards,enoiu2021towards,frasheri2017towards} can be operationalized as argumentative test agents. These agents are intended to capture the features of skilled human testing: they can pursue explicit test goals, adjust their level of autonomy to the situation at hand, interact with other agents and human engineers when needed, and make their reasoning visible through explicit arguments.  This can be operationalized in adaptive test agents that maintain explicit representations of goals, reasons, claims, assumptions, and supporting evidence \cite{Enoiu7368}.

Methodologically, the framework in Figure~\ref{fig:proj} can be operationalized as an iterative loop: collect data and model test-agent behavior, operationalize the relevant constructs, apply the workflow, measure testing efficiency, effectiveness, and reliance outcomes, and use the results to inform the framework. Empirical studies should combine open-source systems with industrial-scale software and use software versions with known naturally occurring faults.

\begin{mdframed}[linewidth=1pt, linecolor=black, roundcorner=5pt]
\small{\textbf{Reliance measures} may include \textit{acceptance of incorrect claims, the frequency of challenged outputs, evidence inspection behavior, human overrides and detection of lost test goals}. The results can be used to iteratively refine the overreliance modes and risks toward an updated model of reliance for test agents.}
\end{mdframed}


\begin{table}[t]
\caption{Overreliance modes and how they relate to AI-assisted software testing.}
\label{tab:overreliance}
\centering
\scriptsize
\begin{tabularx}{\textwidth}{p{0.35\textwidth}p{0.65\textwidth}}
\toprule
\textbf{Mode} & \textbf{Risk Description} \\
\midrule
Test Goal Overreliance & Accepting the test goal selected by the test agent without checking whether it is the right one. \\
\hline
Test Strategy Overreliance & Accepting the chosen technique, e.g., boundary analysis, without asking if another strategy is needed. \\
\hline
Claim Overreliance & Accepting what the test agent says a test demonstrates. \\
\hline
Reason Overreliance & Accepting a plausible reason for why a test agent exists. \\
\hline
Evidence Overreliance & Treating logs, traces, coverage, or summaries as sufficient evidence without checking relevance.  \\
\hline
Test Argument Overreliance & Accepting a coherent generated rationale without challenging the connecting warrant.  \\
\hline
Test Execution Overreliance & Trusting execution summaries instead of inspecting specific failures, skipped tests, uncertainty or evidence. \\
\hline
Oracle Overreliance & Accepting generated expected results or assertions despite weak support.\\
\hline
Delegation Overreliance & Trusting that another test agent handled a delegated subtask correctly. \\
\hline
Regeneration Overreliance & Accepting regenerated tests as improved, equivalent, or sufficient. \\
\hline
Escalation Overreliance & Assuming the test agent will ask for human help when needed. \\
\hline
Maintenance Overreliance & Treating agent-updated tests as preserving the historical test intent(s). \\
\bottomrule
\end{tabularx}
\end{table}

\subsection{Overreliance Modes in AI-Assisted Test Design}
In this section, I apply this proposal to collect data on the conceptual view of test design and test agents. Table~\ref{tab:overreliance} summarizes the modes where overreliance can occur and the risks it can pose in software testing. It shows that overreliance can occur at different points in the tester's problem solving process, from accepting an agent’s test goal or strategy to trusting its claims, evidence, execution, oracles, delegated subtasks, and regeneration. This is intended as an initial checklist for identifying overreliance. This analysis uses a problem-solving lens and a test agent perspective, but offers only a partial view from the author's perspective. Without case study evidence, claims about the overreliance remain interpretive. Nevertheless, overreliance on software testing takes distinct forms when viewed through the lenses of test agents, problem-solving, and test argumentation. 

\begin{mdframed}[linewidth=1pt, linecolor=black, roundcorner=5pt]
\small{\textbf{Illustrative Scenario.} Consider an industrial control system after a requirements change. Test Agent A detects the change and regenerates boundary tests, then delegates input interaction checks to Test Agent B. Agent B reports no failures, but its evidence covers only single interaction cases. Test Agent C flags that an important interaction scenario remains untested. The regenerated tests are assumed to preserve prior intent, limited evidence is treated as sufficient, delegated work is trusted too broadly and a summary hides an untested scenario.}
\end{mdframed}

\section{Conclusions and Limitations}
This paper argues that overreliance on test agents should be understood as an agency, supervisory control, and assurance problem in AI-assisted software testing. By viewing software testing as cognitive problem-solving, we show how agentic AI can shift human work from direct test design to the supervision of test agents. This shift creates testing-specific modes of overreliance, including overreliance on goals, strategies, arguments, oracles, delegation, regeneration, and maintenance. The proposed modes should be treated as a working taxonomy. Future work should study these modes in open-source and industrial settings, develop measures of reliance and quality, and evaluate workflow mechanisms. 

\paragraph{\textbf{Acknowledgments.}}
This work was supported by Software Center, MONA LISA and MATISSE (101140216) projects and the AI and Society Fellowship.

\bibliographystyle{splncs04}
\bibliography{ref.bib}

\end{document}